\documentstyle[12pt]{article}
\begin{document}

\vspace*{0.5cm}
\centerline{PACS no.s 04.60.+n, 03.65.-w, 11.10.-z}
\centerline{Preprint IBR-TH-98-S-023, Dec. 23, 1998}

\vspace*{1cm}
\begin{center}
{\bf UNIVERSALITY OF THE LORENTZ-POINCAR\'E-SANTILLI'S
ISOSYMMETRY FOR THE INVARIANT DESCRIPTION OF 
ALL POSSIBLE  SPACETIMES}
\end{center}

\vskip 0.50 cm
\centerline{{\bf J. V. Kadeisvili}}
\centerline{Institute for Basic Research}
\centerline{P.O.Box 1577, Palm Harbor, FL 34682, U.S.A.}
\centerline{ibr@gte.net; http://home1.gte.net/ibr}

\begin{abstract}
We review the origin of the physical consistency of the Lorentz-
Poincar\'e symmetry. We outline  seemingly 
catastrophic physical inconsistencies
recently identified for noncanonical-nonunitary generalized theories 
defined on conventional spaces over conventional fields. We review
Santilli's isotopic lifting of the Lorentz-Poincar\'e symmetry, by
proving its
invariant resolution of said inconsistencies, and universality for
the representation of all possible spacetimes with a symmetric metric.
The explicit isosymmetry transforms are identified. Particular care is
devoted to the recent discovery of the 11-th dimensionality of
the conventional Poincar\'e symmetry and the consequential emergence of
an axiomatically consistent grand unification of electroweak
and gravitational interactions. The article closes with an outline of
the broader geno- and hyper-symmetries and their isodual for
the description of single-valued irreversible systems,
multivalued irreversible systems and antimatter systems, respectively.
\end{abstract}

\vskip 0.50 cm

\noindent {\large \bf 1. Lorentz-Poincar\'e Symmetry}.

\noindent Physics is a discipline admitting the reduction of
events to primitive symmetries, the most important ones being the
symmetries of our spacetime [1], namely,
the rotation, boosts, translation and discrete symmetries,
hereon called the {\it Lorentz -Poincar\'e symmetry}
(or the {\it L-P symmetry} for short) and denoted P(3.1).

We are referring to the most general possible,
linear, local-differential and canonical (for classical formulations)
or unitary (for operator formulations)
symmetries of the Minkowski space $M = M(x,\eta,R)$ with:
spacetime coordinates $x = \{x^\mu\} = (x^k, x^4), x^4 = c_ot,
\mu = 1, 2, 3, 4, k = 1, 2, 3,$ $c_o$ being the speed of light in
vacuum;
unit I = Diag. (1, 1, 1, 1);
 metric $\eta = Diag. (1, 1, 1, -1)$; and
invariant on the field $R = R(n,+,\times)$ of real numbers n with
conventional sum + and associative product $\times$
$$
(x - y)^2 = (x - y)^{\mu}\times \eta_{\mu\nu}\times (x - y)^\nu =
$$
$$
= (x-y)^1\times (x-y)^1 + (x-y)^2\times (x-y)^2 + (x-y)^3\times (x-y)^3
-
(x - y)^4\times (x-y)^4 = inv.
\eqno(1.1)
$$

All spacetime theories possessing the Lorentz-
Poincar\'e symmetry have an impeccable axiomatic and physical
consistency, as it is the case for
relativistic quantum mechanics, special relativity, unified
gauge theories of electroweak interactions, and other theories.

These historical successes of the L-P symmetry are due to the {\it
invariant}
(rather than covariant) character
of the theories, which, in turn,
is permitted by their (canonical or) unitary structure
on a Hilbert space $\cal H$ over the field $C(c,+,\times)$
of complex numbers c,
$$
 U\times U^{\dagger} = U^{\dagger}\times U = I.
\eqno (1.2)
$$

The fundamental Lorentz-Poincar\'e invariance
begins with the invariance under the time
evolution of the theories, and implies
the numerical invariance of the basic units used for measurements,
the preservation in time of Hermiticity-observability, the
invariance of the special functions and transforms used in data
elaboration,
the uniqueness and invariance of the numerical predictions, and other
features essential for physical consistency.

In the final analysis, the above  mathematical and physical consistency
can be
traced to the fact that
classical or operator Lorentz-Poincar\'e invariant theories
possess a {\it Lie structure.}

Even though well known, it may be useful for subsequent referrals to
recall the basic invariances for unitary theories
$$
I\rightarrow U\times I\times U^{\dagger} = I' = I,
$$
$$
A\times B\rightarrow U\times (A\times B)\times U^{\dagger} =
U\times A\times U^{\dagger}\times U\times B\times U^{\dagger} = A'\times
B',
$$
$$
H\times |\psi > = E\times |\psi>\rightarrow U\times H\times |\psi > =
U\times H\times U^{\dagger}\times U |\psi >
= H'\times |{\psi}'> =
$$
$$
U\times E\times |\psi> = E'\times |{\psi}'>, E' = E,
\eqno (1.3)
$$
\vskip 0.50 cm
{\it THEOREM 1: All theories with a unitary structure on a Hilbert space
over the field of complex numbers possess numerically invariant units,
products and
eigenvalues, thus being suitable to represent physical reality}.
\vskip 0.50 cm

\noindent {\large \bf 2. Inconsistencies of
Noncanonical-Nonunitary Generalizations.}

\noindent This paper will have achieved its first objective if it
contributes to stimulate the awareness by the
contemporary physica community to come to its senses, and
address the rather serious physical
 inconsistencies of  spacetime theories with a noncanonical or
nonunitary structure treated via the mathematics of
canonical or unitary theories.

Physics is a quantitative
science in which, sooner or later, physical truths always emerge.
Therefore, silence on these inconsistencies
can only damage the authors of papers on noncanonical-
nonunitary theories.

The lack of universality of the Poincar\'e symmetry for the
description of the entire universe was identified immediately
following its appearance and then confirmed throughout this century.
This scientific process lead to the construction of numerous theories
representing events in our spacetime, yet violating the
Lorentz-Poincar\'e
axioms in favor of broader axioms.

No understanding of the topic of this paper
(the isotopies of Lorentz-Poincar\'e) can be claimed without at
least a rudimentary knowledge of the now considerable literature
on the indicated inconsistencies.

The first generalization is due to Einstein himself who, immediately
following the formulation of the special relativity, identified the
impossibility of representing gravitation with the realization of the
Lorentz-Poincar\'e
axioms of the time, and formulated the general theory of relativity
on Riemannian spaces [2].

While Einstein's studies based on the Lorentz-Poincar\'e symmetry have
passed the test of time and are nowadays more valid than ever,
Einstein's
theory of gravitation, which departs from said symmetry,
has been the subject of endless, still
unresolved and actually increasing
controversies during this century (see, e.g., representative
papers [3] and references quoted therein).

The origin of most of these controversies has been recently identified
by Santilli [3f] and can be summarized as follows. The map
from the Minkowski metric $\eta$ to
the Riemannian metric g(x) is clearly a {\it noncanonical}
transformation at the
classical level and a {\it nonunitary transformation} at the operator
level,
$$
\eta\rightarrow g(x) = U(x)\times \eta\times U^{\dagger}(x), U\times
U^{\dagger}
\not = I.
\eqno (2.1)
$$.

As a result, {\it any theory on a curved manifold
is structurally noncanonical-nonunitary, beginning with its time
evolution.}

Despite an undeniable {\it mathematical beauty} that has
attracted so many scholars
throughout this century,
a host of rather serious problems of
{\it physical consistency} then follows.
\vskip 0.50 cm

{\it THEOREM 2 [3f]: All theories with a nonunitary structure
on a conventional Hilbert space
over the field of complex numbers,
thus including (but not limiting to) all
operator theories of gravity formulated on a curved manifold, possess
the
following physical inconsistencies:

        1) lack of invariant units of space,
time, energy, etc.,
 with consequentially impossible
applications to real measurements;

2) lack of preservation of the original Hermiticity in time, with
consequential
absence of physically acceptable observables;

3) general violation of causality and
probability laws;

4) lack of invariance of conventional and special
functions and transforms used in data elaborations;

5) lack of uniqueness and invariance of numerical predictions;

and have other inconsistencies which render them inapplicable
to represent physical reality}
\vskip 0.50 cm

The proof of these occurrences is elementary. The lack of invariance of
the
basic units is inherent in the very conception of nonunitary transforms
(see later on for details).
The lack of preservation in time of Hermiticity-observability is known
as {\it Lopez's lemma} [3g]. The violation of probability laws is an
evident
consequence of the lack of invariance of the basic units, with
consequential violation of causality. Nonunitary transforms do not
preserve elementary functions such as the exponentiation, let alone
special functions and transforms. The lack of uniqueness of
the numerical predictions is evident from the lack of uniqueness of
the value of nonunitary transforms, while the lack of
invariance of the numerical predictions is so evident to
require no comments.

Even though known, it may have graphical value to review the
fundamental noninvariances under nonunitary transforms from which
all the physical inconsistencies follow [3f]:
$$
I\rightarrow U\times I\times U^{\dagger} = I' \not = I,
$$
$$
A\times B\rightarrow U\times (A\times B)\times U^{\dagger} =
 U\times A\times U^{\dagger}\times (U\times
U^{\dagger})^{-1}\times U\times B\times U^{\dagger} =
A'\times T\times B', T = (U\times U^{\dagger})^{-1},
$$
$$
H\times |\psi > = E\times |\psi>\rightarrow U\times H\times |\psi > =
U\times H\times U^{\dagger}\times (U\times U^{\dagger})^{-1}\times U
\times |\psi >
=
$$
$$
H'\times T\times |{\psi}'> =
U\times E\times |\psi> = E'\times |{\psi}'>, E' \not = E,
\eqno (2.2)
$$
namely, {\it under nonunitary time evolutions and
transforms we have the alteration
of the numerical value of all basic units, all product and all
eigenvalues}.

Santilli [3f,5l,6c] has identified additional catastrophic
inconsistencies which apply to both
noncanonical and nonunitary theories. Recall that all physical
theories are based on numbers and fields which, in turn,
are based on the fundamental (multiplicative) unit. The alteration of
the
unit by noncanonical-nonunitary transforms then implies the shift to
{\it different numbers and fields}.
But noncanonical-nonunitary theories continue to be defined on
{\it conventional} numbers and fields. This implies the collapse of the
axiomatic consistency of the entire theory, including the
inapplicability of vector and metric spaces, geometries and topologies,
algebras, groups and symmetries, etc., with no known exceptions.

Note that the latter arguments rules out the physical value of any
{\it classical} noncanonical theories, again, because they imply the
alteration of the basic unit with consequential
inapplicability of the carrier spaces used to elaborate the theory.

The above catastrophic physical and axiomatic inconsistencies apply in
their
entirety to the classical and operator formulation of gravity on
curved manifolds. As an example, there is no known physical meaning or
consistency in attempting the "experimental
verification" of the general relativity at a given time t
defined via field equations
on a Riemannian space over the fields of real numbers, when the basic
unit
I is altered at a subsequent time t', Eq. (2.2a). We then have the
consequential lack of physical
meaning in preserving the Riemannian space itself
because defined on a field no longer applicable at t'.
The physical inconsistencies of the operator
formulation of gravitation on a curved space are so
serious and transparent to require no further comments.

The ultimate origin of the above gloomy scenario investing
about one century of studies in gravitation
is the very notion of {\it curvature}
itself, because it implies the {\it breaking of the
fundamental Lorentz-Poincar\'e symmetry} in favor of "covariance"
under a broader, often undefined symmetry, with the indicated
catastrophic consequences. In fact, the Lorentz-Poincar\'e invariance
and the notion of curvature
are mutually exclusive in a transparent and irreconcilable way
in their current formulation (see next sections for an alternative).

The limitations of the Lorentz-Poincar\'e symmetry
have also been felt by numerous other scholars besides Einstein,
particularly during the
recent decades. We here quote: the studies by Y. S. Kim
and others (see [4a] and references quoted therein), which have
the important function of extending
the applicability of the Lorentz-Poincar\'e axioms to their ultima
possibilities
for the representation of extended particles; the use of broader
symmetries
in an attempt to reach a grand unification inclusive of the
gravitational
interactions (see, e.g., [4b]); the broadening of the Lorentz-Poincar\'e
symmetry inherent in contemporary string theories [4c]; and
numerous other theories (see other papers in this collection [4d]).

It is important for the contemporary physics community to study,
understand
and, above all, admit that {\it all generalized theories
with a noncanonical or nonunitary structure,
even though possessing an undeniable mathematical beauty,
have no known physical application}.

Along these lines, in memoir [6e] of 1996, Santilli clearly states
the physical inconsistency of his
{\it Birkhoffian generalization of Hamiltonian mechanics} published
in monograph [6g] (by Springer-Verlag in its most prestigious
physics series...), precisely because of its noncanonical structure
formulated
on conventional spaces over conventional fields. The reader should be
aware that
the Birkhoffian  mechanics was proved in the same monograph to be
universal for
all well behaved, local-differential and nonhamiltonian
systems with a
generalized Lie-isotopic structure.
In the same memoir [6e] Santilli clearly states
the additional; physical inconsistency of his broader
classical Lie-admissible mechanics of monograph
[12f] which is universal for
all Newtonian system with a non-Lie, yet algebraically consistent
structure.
In the same memoir [6e] Santilli
presents new invariant classical mechanics of Lie-isotopic and
Lie-admissible
type
we cannot possibly review here for brevity.

Similarly, in memoir [5l] of 1997, Santilli
clearly states the physical inconsistency of {\it all} his
generalized operator studies prior to 1997, including all numerous
papers written on hadronic mechanics since its proposal of 1978 [6b],
including all papers on operator Lie-isotopic and Lie-admissible
theories
(which are also universal for all possible nonlinear, nonlocal and
nonunitary
theories with and without an antisymmetric algebras, respectively).
In the same memoir [5l]
Santilli proposes fully invariant operator, Lie-isotopic
and Lie-admissible formulations we shall
outline in the next sections).

Regrettably, the same clear statements of physical inconsistencies are
lacking at this writing, to our best knowledge, on numerous
other generalized theories with a transparent
and incontrovertible  nonunitary structure,
each theory possessing a rather vast literature, such as
(see [3f] for complete list and references):

1) Dissipative nuclear models with imaginary potentials,
$H = H_o + iV$, and time evolution
$idA/dt = A\times H^{\dagger} - H\times A = [A, H, H^{\dagger}]$
(these theories
lose an "algebra" as commonly understand, in favor of a triple
system - as a result of which names such as "proton" and "neutron"
lose their physical meaning because of the impossibility
to even define spin, mass and other basic characteristics,
let alone treat them);

2) Statistical models  with external collisions terms
with time evolution $ id\rho/dt = [\rho, H] + C$ (besides being
nonunitary, these theories have no units al all - let alone
a noninvariant units - and have no exponentiation at all,
under which catastrophic conditions any application to
physical reality implies exiting science);

3) q- deformations of the Lie product $A\times B - q\times B\times A$,
"*-deformations" of the enveloping associative algebra
'with generalized product $A*B = A\times T\times B$, and
other deformations which change
the Lie structure while preserving the old mathematics, all being
transparently nonunitary (all these deformations were
first introduced by Santilli in his Ph. D.
Thesis of 1967 [12a], although this paternity is
ignored in the rather vast literature in the field,
evidently to the sole detriment of the authors);

4) Certain quantum groups (evidently those
with a nonunitary structure);

5) Weinberg's nonlinear theory with
nonassociative Lie-admissible envelopes (which lacks any unit,
violates Okubo's no quantization theorem prohibiting
the use of nonassociative envelopes [3h], and has other serious flaws);

6) All known theories of quantum gravity (the indication
of theories in this field with a unitary structure would be
appreciated);

7) All known supersymmetric theories (evidently because they broaden
the very structure of Lie algebras and groups via the
addition of anticommutators, thus resulting in an evident nonunitary
structure);

8) all known studies on Kac-Moody superalgebras (also because
they depart from Lie's structure with a phase
term depending on anticommutators);

9) All known string theories whose nonunitary structure was known
since the introduction of the Beta function by
Veneziano and Suzuki, and reinforced via supersymmetries in the recent
studies (see the specific study [3i]).

Other theories which have a seemingly unitary structure, but depart from
other axioms of Lie's theory equally possess serious physical flaws.
This is the case, for instance
for theories with Hermitean Hamiltonians, yet a structure
{\it nonlinear in the wavefunction} of the type $H(x,p,\psi,..)\times
|\psi > = E\times |\psi >$
 (again, see [3f] for details and regferences).
These theories violate the superposition principle,
thus being inapplicable to composite systems; they violate Mackay
imprimitivity
theorem, thus violating the integrability conditions for
the Galilean and Einsteinian relativities; and have other other flaws.

Yet other theories violate the {\it locality} condition of Lie's theory,
e.g., via "integral potentials" in the Hamiltonians. These theories
are fundamentally flawed on both mathematical grounds (because the
assumption is incompatible with the basic topology)
and physical grounds (because
nonlocal interactions generally are of contact-zero range type, thus
having no potential). As such, these theories deserve no further comment
(or attention).

In summary, Santilli has established that {\bf \it all theories which
violate any of the
fundamental axioms of linearity, locality and canonicity-unitarity of
Lie's theory is  physically inconsistent when formulated
via the mathematics of quantum mechanics}.

In other cases, the existence of possible inconsistencies
requires specific investigations. This is the case of Kim's [4a] theory
which replaces the Lorentz-Poincar\'e {\it invariance} with a broader
{\it covariance}. These studies are left to the interested readers.

We close this section by indicating that {\it classical} theories of
{\it antimatter} are generally inconsistent
because they only have one channel of
quantization for matter and antimatter. As a result, their
orator image {\it does not} yield
charge conjugate states, but merely states of particle
with the wrong sign of the charge.

The Riemannian treatment of antimatter is afflicted by more catastrophic
physical inconsistencies because, in addition to the above
inconsistent operator image, they can only represent antimatter via
the usual energy-momentum tensors which are
 notoriously {\it positive-definite}, thus being in dramatic
disagreement
with the {\it negative-definite} energies need for antiparticles.

These inconsistencies should not be surprising because
the biggest unbalance in the physics
literature of this century is precisely the
treatment of matter at all possible
levels, from Newton to quantum field theory, while antimatter
is solely treated at the level of second quantization. But
antimatter is expected to exist at the macroscopic level, i.e.,
that of entire galaxies or quasars, thus demanding the restoration
of a fully equivalent treatment of matter and antimatter at
all levels of study.

By no means all generalized theories of the contemporary
physical literature are wrong. In fact, numerous generalized theories
constructed on sound foundations have an impeccable axiomatic structure,
such as the
theories by  Ahluwalia [4e],  Dvoeglazov [4f], and others.

\vskip 0.50 cm

\noindent {\large \bf 3. Lorentz-Poincar\'e-Santilli isosymmetry}

\noindent By initially working in a rather solitary way,
the Italian-American physicist R. M. Santilli [5]  has constructed a
new realization of the Lorentz-Poincar\'e axioms which:

1) is "directly universal" for the representation of all infinitely
possible, nonlinear, nonlocal and noncanonical-nonunitary
theories in our (3+1)-dimensional spacetime with a
well behaved, nowhere singular and symmetric metric (universality),
directly in the x-coordinates of the observer without any
use of the transformation theory (direct universality);

2) reconstructs the canonicity or unitarity and invariance, on
suitably generalized spaces over generalized fields; and

3) resolves the physical inconsistencies indicated in Sect. 2.

Remarkably,
Santilli [5] constructed the most general known symmetry of the
following most general possible invariant in (3+1)-dimensions with
the indicated topological condition on the metric:
$$
(x - y)^{\hat 2} = (x - y)^\mu\times \hat {\eta}_{\mu\nu}(x, v, d, \tau,
\psi, ...)
\times (x - y)^\nu =  (x - y)^\mu\times
\hat T_{\mu}^{\rho}(x, v, d, \tau, \psi, . .)\times
{\eta}_{\rho\nu}\times (x - y)^\nu =
$$
$$
=(x-y)^1\times \hat T_{11}(x, v, d, \tau, \psi,...)\times (x-y)^1 +
(x-y)^2\times \hat T_{22}(x, v, d, \tau, \psi, ...)\times (x-y)^2 +
$$
$$
+ (x-y)^3\times \hat T_{33}(x, v, d, \tau, \psi, ...)\times (x-y)^3 -
(x - y)^4\times \hat T_{44}(x, v, d, \tau, \psi, ...)\times (x-y)^4 =
inv.
\eqno(3.1)
$$
where all functions $\hat T_{\mu\nu} (= \hat T_\mu^\nu)$
are positive definite but
otherwise possess an unrestricted, generally {\it nonlinear,
non local and nonhamiltonian} functional dependence on
spacetime coordinates
x, velocities v, density d, temperature $\tau$, wavefunctions $\psi$, or
any other needed local quantity.

Unexpectedly, Refs. [5] then proved that
the universal symmetry of interval (3.1) is locally isomorphic
to the symmetry of the {\it conventional} invariant (1.1),
of course, when properly formulated. In fact, Santilli
insists in his writings
that the symmetry of invariant (3.1) {\it is not new}, because it is
merely a
{\it new realization} of the conventional Lorentz-Poincar\'e axioms.
This implied the reconstruction of the
Lorentz-Poincar\'e symmetry as being {\it exact} when popularly believed
to be
broken, as we shall see (e.g., for gravitation).

Santilli [5] then proved the "direct universality" of this symmetry via
the explicit construction of the most salient applications.

These results were achieved via the prior construction of a new
mathematics, originally proposed in Ref. [6a] under the
name of {\it isomathematics} from the
Greek meaning oif being "axiom-preserving", and then developed by
various authors [6-8] (see [7n]
for a comprehensive literature up to 1984, [5o] for literature up to
1995,
and Web Site [7o], Page 18, for a readable outline).
The new mathematics is essentially characterized by new numbers, new
fields,
new spaces, new algebras, etc. called
isonumbers, isofields, isospaces, isoalgebras, etc.
For this reason the universal
symmetry of invariant (3.1) is known as the
{\it Lorentz-Poincar\'e-Santilli isosymmetry} (also called the
{\it L-P-S isosymmetry} or {\it Santilli's
isopoincar\'e symmetry} for short),
and it is generally denoted $\hat P(3.1)$ [6-9].

The main working ideas are essentially the following:

1) the generalization (called {\it lifting}) of the Minkowski metric
$\eta$ into the most general possible, well behaved,
nowhere singular and symmetric
metric $\hat {\eta}(x, v, d, \tau, \psi, ...) =
\hat T(x, \-v, \-d, \tau, \psi, ...)\times \- \eta$, where $\hat T$ is
a $4\times 4$ well behaved, nowhere singular and {\it positive-definite}
(thus diagonalizable) matrix;

2)  the joint lifting of the fundamental unit of the Minkowski space,
I = Diag. (1, 1, 1, 1), by the {\it inverse} of the lifting of the
metric, $\hat I = 1/\hat T$; and

3) the reconstruction of the entire mathematical
foundations of Lorentz and Poincar\'e into a
form admitting $\hat I$, rather than I, as the correct left and
right unit of the new theory.

The latter condition requires the lifting of the
conventional associative product $A\times B$ among generic quantities
A, B (numbers, matrices, operators, etc.) into the form $A\hat {\times}
B =
A\times \hat T\times B$, with $\hat T$ fixed, for which
$\hat I\hat {\times} A = A\hat {\times} \hat I = A$ for all possible A.
In this case (only), $\hat I$ is called the {\it isounit},
and $\hat T$ is called the {\it isotopic element}.

In turn, the latter liftings imply, for evident reason of consistency,
the new {\it isofields}
$\hat R = \hat R(\hat n,\hat +, \hat {\times})$ [6b] of
{\it isonumbers} $\hat n = n\times \hat I$ with
{\it isosum} $\hat n\hat +\hat m = (n+m)\times \hat I$,
{\it isoproduct} $\hat n\hat {\times}\hat m = (n\times m)\times \hat I$,
{\it isoquotient} $A\hat {/}B =
(A/B)\times  \hat I$, and other generalized operations.

Under the above conditions,
it is evident that $\hat R$ and R are isomorphic, and actually coincide
at the
abstract level (because $\hat I$ and $I$ are topologically identical).
Despite this simplicity,
the reader should abstain from jumping at conclusion of mathematical
triviality to avoid insidious misrepresentations. As an illustration,
"two multiplied by two is sixteen" and
the number 4 becomes {\it prime} for isounit $\hat I =$ 4. This
indicates the
dependence of number theory from the assumed unit. Following
memoir [d] the {\it Santilli's isonumber
theory} has been the subject of comprehensive studies by C. X. Jiang
[7g,7m],
Kamiya [7h], Trell [7i], and other mathematicians.

By recalling that metric spaces are defined on a given field,
the availability of new numbers and fields permitted the
construction of the isotopies of the Minkowski space,
presented for the first time in Ref. [5a] (see also [5,6]), today
called
{\it Minkowski-Santilli isospaces}
and denoted $\hat M = \hat M(\hat x,\hat {\eta}.\hat R)$
with {\it spacetime isocoordinates} $\hat x = x\times \hat I$ defined
precisely on $\hat R$, and consequential lifting of algebras, groups,
geometries, topologies, etc. [5,6,7].

Under the above liftings, i.e.,
$$
\eta\rightarrow \hat {N} = (\hat {\eta}_{\mu\nu}\times \hat I)
= (\hat T_{\mu}^{\rho}\times \eta)_{\rho\nu}\times \hat I, \hat T > 0,
I\rightarrow \hat I = 1/\hat T, A\times B \rightarrow A\hat {\times} B =
A\times \hat T\times B, etc.
\eqno (3.2)
$$
the new isospaces $\hat M$ are locally
isomorphic to the conventional space $M$; the
isosymmetry $\hat P(3.1)$ is locally isomorphic to the
conventional symmetry P(3.1); and  {\it all} properties, axioms and
physical laws holding on M over R admit an {\it
identical} image on $\hat M$ over $\hat R$. These are the reasons
for the original suggestion of the name
{\it isotopies} [6a] from the Greek meaning of being "axiom-preserving".

In this way, the isorelativistic theories {\it coincide}, by conception
and construction, with conventional relativistic theories at the
abstract,
realization-free level, by therefore bringing the applicability
of the Lorentz-Poincar\'e symmetry and Einstein special relativity
to the unexpected level of universality.

Moreover, Santilli [5] proved that {\it the conventional Poincar\'e
symmetry is eleven
dimensional, and not ten dimensional
as believed throughout this century}. This additional unexpected
property
was proved via the
new invariance of the Minkowskian line element [6e],
$$
(x^\mu\times \eta_{\mu\nu}\times x^\nu )\times I =
[x^\mu\times ({\rho}^{-2}\times\eta_{\mu\nu})\times x^\nu )\times
({\rho}^2\times I) =
(x^\mu\times \hat {\eta}_{\mu\nu}\times x^\nu )\times \hat I,
\eqno (3.3)
$$
where $\rho$ is an ordinary parameter, with corresponding novel
invariance
of the Hilbert product [5j]
$$
<\phi |\times |\psi >\times I =
<\phi |\times {\rho}^{-2}\times |\psi >\times ({\rho}^2\times I) =
<\phi |\hat {\times} |\psi >\times \hat I.
\eqno (3.4)
$$

It is evident that Eqs. (3.3) characterizes the isominkowski spaces
$\hat M$ over $\hat R$ in their simplest possible realization, that with
isounit characterized by an ordinary parameter,
$\hat I = {\rho}^2$. Eqs. (3.4) then
characterize the
simplest possible realization of the
{\it isohilbert spaces} $\hat {\cal H}$
defined on the isofield $C(\hat c,\hat +,\hat {\times})$
of isocomplex numbers $\hat c = c\times \hat I$, used for the operator
formulation of the isosymmetry. It is also evident that
the above new symmetries persists at the full isotopic level,
$$
(x^\mu\times \hat {\eta}_{\mu\nu}\times x^\nu )\times \hat I =
[x^\mu\times ({\rho}^{-2}\times \hat {\eta}_{\mu\nu})\times x^\nu
]\times
({\rho}^2\times \hat I) =
(x^\mu\times \hat {\eta}'_{\mu\nu}\times x^\nu )\times \hat {I}',
$$
$$
<\phi |\hat {\times} |\psi >\times \hat I =
<\phi |\times {\rho}^{-2}\times \hat T\times |\psi >\times
({\rho}^2\times \hat I) =
<\phi |\hat {\times}' |\psi >\times \hat {I}'.
\eqno (3.5)
$$
As a result,
the {\it Lorentz-Poincar\'e-Santilli isosymmetry is also eleven
dimensional }
(see Sect. 5 for details).

By recalling that any new symmetry of spacetime has far reaching
physical
implications, Santilli's discovery of a hitherto unknown additional
dimension of the fundamental symmetries of our spacetime
also has important and novel physical
implications outlined below.

The reader should not be surprised that the new symmetry (3.3) has
remained
unknown since Lorentz-Poincar\'e-Minkowski's times,
and the additional new symmetry (3.4)
has remained unknown 'since Hilbert's time. In fact,
their identification required
the prior
discovery of {\it new numbers}, those {\it with arbitrary units} [6d].

As a guide to the existing main literature, we here indicate the
first construction of the isotopies of:
rotational symmetry in Ref. [5b]; Lorentz symmetry in Ref [5a];
SU(2)-spin symmetry in Refs. [5c,5d]; Poincar\'e symmetry
in Ref. [5e]; and spinorial covering of the
 Poincar\'e symmetry in Ref. [5f]. In Refs. [5g,5h]
Santilli achieved the first axiomatically consistent grand
unification of electroweak and gravitational interactions known to this
author
precisely via the use of the 11-dimensional isopoincar\'e symmetry;
and in Ref. [5i] he presented
the isopoincar\'e  invariant {\it isocosmology.}
In memoir [5j] one can find
a comprehensive presentation of the underlying isominkowskian
geometry and related reformulation of gravity;  the operator
formulations originated in paper [6b] (of 1978), continued in numerous
publications (see, e.g., Ref. [5k,5-1o,12-14]), and reached maturity in
memoir [5l].
Classical realizations of the (isogalilean and)
isopoincar\'e symmetries were studied in detail
in monographs [5m,5n], while the operator counterparts were
studied in detail
in monographs [5o,5p].

Pre-requisites for the above results were the isotopies of Lie's
theory in its various branches,
the universal enveloping associative algebras (including the
Poincar\'e-Birkhoff-Witt theorem), Lie algebras (including the
celebrated
Lie first, second and third theorem), Lie's groups,
transformation and representation theories. These isotopies were
proposed for the first time in Ref. [6a], and then studied in a
variety of works (see monograph [6g] for the status of the knowledge as
of 1983,
and monograph [5o] for the status as of 1995). The emerging theory is
today properly called {\it Lie-Santilli isotheory} and it is
the subject of numerous independent studies, such as those by:
Tsagas and Sourlas in the papers of Refs. [7] and monograph [7j];
Lohmus,  Paal and Sorgsepp in monograph [7k]; Vacaru in papers
[7] and monograph [7l]; Kadeisvili in Refs. [8]; and
additional authors quoted therein (see the miscellaneous list of papers
[9]).

It is evident that we cannot possibly
provide a technical treatment in this note
of all the above results. To avoid lecture-notes
for a two-semesters course, we must, therefore,
restrict ourselves to only the most essential aspects.

The feature of paramount importance for these introductory lines is the
reconstruction on isohilbert spaces $\hat {\cal H}$ over isofields
$\hat C(\hat c,\hat +\hat {\times})$ of unitarity for
all conventionally nonunitary transforms, according to the
{\it isounitarity conditions}
$$
\hat U\hat {\times} \hat U^{\dagger} = \hat U^{\dagger}\hat {\times} U =
\hat I,
\eqno (3.6)
$$

In particular, all possible conventionally nonunitary transforms
on $\cal H$ over C can always be
identically rewritten in the isounitary form on $\hat {\cal H}$ over
$\hat C$
(first identified in [5l])
$$
U\times U^{\dagger} \not = I, U = \hat U\times \hat T^{1/2},
\eqno (3.7)
$$

Once such an isounitary structure is achieved,
it remains invariant under all possible, additional isounitary
transforms,
$$
\hat W\times \hat W^{\dagger} = \hat W^{\dagger}\times \hat W = \hat I,
$$
$$
\hat I\rightarrow \hat W\hat {\times}\hat I\hat {\times} \hat
W^{\dagger}
= \hat I,
$$
$$
\hat A\hat {\times} \hat B\rightarrow \hat W\hat {\times}
(\hat A\hat {\times}\hat B)\hat {\times}
\hat W^{\dagger} = \hat {A}'\hat {\times}\hat {B}',
$$
$$
\hat H\hat {\times} |\hat {\psi}> = E\times |\hat {\psi}>\rightarrow
\hat W\hat {\times}\hat H\hat {\times} |\hat {\psi> =
\hat {H}'\hat {\times} |\hat {\psi}'> =
\hat W\hat {\times}\hat E\hat {\times} |\hat \psi}> =
 = E\times |\hat {\psi}'>
\eqno (3.8)
$$

Note {\it the invariance of the numerical values of the isounit,
isoproduct and the
isoeigenvalues}, as necessary for physical consistency. Classical
noncanonical transforms are similarly turned into {\it identical}
isocanonical
versions with resulting invariance not considered here for brevity.

In summary, all nonunitary transforms are rewritten in an
{\it identical} isounitary form which reproduces all the
original invariances of conventionally
unitary theories, thus resolving the inconsistencies of Sect. 2.

Along the same lines, Santilli reconstructs theories
that are nonlinear (in the wavefunction) on $\cal H$ over C into
identical {\it isolinear} forms on $\hat {\cal H}$ over $\hat C$
via the identifications $H(r, p, \psi)\times |\psi> =
H_o(r, p) \hat T(\psi, ...)\times |\psi> =
H_o(r, p)\hat {\times} |\psi> = E\times |\psi>$,
namely, by embedding all nonlinear terms in the isotopic element. This
reformulation implies the regaining of the superposition principle,
and the resolution of the other inconsistencies.

Similarly,isotheories are {\it nonlocal-integral}
(e.g., because admitting volume integrals to represent
wave-overlappings). These theory are however reconstructed
as local-differential
on isospaces over isofields, called {\it isolocal-isodifferential},
via the embedding of all nonlocal terms in the isotopic element.

In this way, the universal symmetry of invariant
(3.1) is the largest possible isolinear, isolocal and isocanonical or
isounitary symmetry of isospacetime.

     In inspecting the literature on isotopies, the reader should keep
in mind
that all references prior to memoirs [5l,6e], even though
formulated on isospaces over isofields, {\it are not invariant}.
After laborious
studies, Santilli identified the origin of the
problem where one would expect it the least,
in the {\it ordinary differential calculus} which, contrary
to popular beliefs, resulted to be dependent on the fundamental unit
of the base field. This point is absent on
the vast literature on different calculus
through various centuries because for the
tacitly assumed trivial unit I = +1, we
have d(+1) = 0, while for more general units with a
nontrivial functional dependence, we
evidently have $d\hat I(x, v, ...0 \not = 0$.

The
latter occurrence required a reformulation
of the differential calculus into a form, called
{\it isodifferential calculus,} which is compatible
with the generalized unit of the base field, first
achieved by Santilli in memoir [6e] via the main rules
$$
\hat d\hat x^\mu = \hat I^{\mu}_{\nu}\times d\hat x^\nu,
\hat {\partial} / \hat {\partial} \hat x^\mu =
\hat T_{\mu}^{\nu}\times \partial / \partial \hat x^{\nu},
\hat {\partial}\hat x^{\mu}\hat /\hat {\partial}\hat x^{\nu} =
\hat {\delta}^{\mu}_{\nu} = \delta^{\mu}_{\nu}\times \hat I.
\eqno(3.9)
$$

The above new calculus was then applied in memoir [5j] to the
construction
of a novel geometry, the {\it isominkowskian geometry}
which resulted to be a symbiotic unification of the Minkowskian
features (as reported above), plus the machinery of Riemann
 (because the isominkowskian metric has an x-dependence),
including the
 {\it isochristoffel's symbols,  isocovariant isodifferential,
 isocovariant isoderivative,} etc.,

{\it isocurvature tensor}
$$
\hat {\Gamma} _{\alpha\beta\gamma} =
\hat{\frac{1}{2}}\hat\times  (\hat{\partial}   _{\alpha    }\hat{\eta
}_{\beta\gamma } +
\hat{\partial }_{\gamma }\hat{\eta }_{\alpha\beta } -
\hat{\partial }_{\beta }\hat{\eta }_{\alpha\gamma })\times \hat{I},
\hat{D}\hat{X}^{\beta} = \hat{d}\hat{X}^{\beta} +
\hat{\Gamma }_{\alpha }^{\beta }{}_{\gamma }\hat {\times}
\hat{X}^{\alpha }\hat {\times} \hat{d}\hat{x}^{\gamma }
$$
$$,
\hat{X}^{\beta}_{|\hat{}\mu} =
\hat{\partial }_{\mu }\hat{X}^{\beta} +
\hat{\Gamma }_{\alpha }^{\beta }{}_{\mu } \hat {\times} \hat{X}^{\alpha
},
\hat{R}_{\alpha }^{\beta }{}_{\gamma\delta }
=\hat{\partial }_{\beta }\hat{\Gamma }_{\alpha }^{\beta }{}_{\gamma } -
\hat{\partial }_{\gamma }\hat{\Gamma }_{\alpha }^{\beta }{}_{\delta } +
\hat{\Gamma }_{p}^{\beta }{}_{\delta } \hat {\times} \hat{\Gamma
}_{\alpha }^
{p}{}_{\gamma } -
\hat{\Gamma }_{p}^{\beta }{}_{\gamma } \hat {\times} \hat{\Gamma
}_{\alpha }^
{p}{}_{\delta }.
\eqno (3.10)
$$

The isominkowskian geometry then permitted the {\it identical}
formulation
of conventional gravitational field equations, such as the
Einstein-Hilbert field equations, althougfh now formulated in a space
which is {\it isoflat}, thus resolving the main problems
of the conventional formulation outlined in Sect. 2 (see Sect. 4 for
details).

By keeping in mind that conventional and isotopic
differentials and derivatives coincide at the abstract
level, all papers on isotopies prior to 1996 can be easily
completed into a fully invariant form via the mere re-interpretation of
the symbols "d" and $"\partial"$ as being isotopic.

Numerous applications and experimental verifications of the
isorelativistic theories have been worked out to date
by various authors, among which we indicated:

{\bf A) Particle physics}: the universality of the isominkowskian
geometry
for the geometrization of all physical media,
whether of low density (such as our atmosphere) or of high density (such
as the interior
of hadrons and stars) with an excellent fit of experimental data [10a];
the universality of the Lorentz-Poincar\'e-Santilli
isosymmetry for the representation of arbitrary local speeds of light
[10b]
as established by evidence [11]; the
exact-numerical representation of the
Minkowskian anomalies in the behavior of the meanlife of unstable
hadrons
with speed [10c]; the exact-numerical representation
of the experimental data from
the Bose-Einstein correlation [10d]; the achievement of a true
confinement of
quarks (with an identically null probability of tunnel
effects for free quarks) within
a perturbatively convergent theory and conventional
SU(3) quantum numbers [10e,10f,10g];
the reconstruction of the {\it exact parity, Lorentz and Poincar\'e
symmetries} in particle physics when believed to be broken [5p];
and other verifications.

{\bf B) Nuclear Physics}: the reconstruction of
the {\it exact isospin symmetry} in nuclear physics
[5d]; the first exact-numerical representation of {\it all}
total nuclear magnetic moments via the invariant representation of
the deformation of shape of the nucleons [10h];
the first exact representation of the synthesis of neutrons
as they occur in stars at their beginning, from protons and electrons
{\it only} (thus excluding the yet unavailable remaining baryons with
consequential impossibility to use quark theories) [5f]; the
prediction that the neutron, a {\it naturally unstable} particle,
can be {\it stimulated to decay} via suitable resonating mechanisms
which are possible for a nonunitary theory although simply inconceivable
for
quantum mechanics, and consequential prediction
of a novel {\it subnuclear} energy currently under
industrial development [10i]; and other verifications.

{\bf C) Astrophysics and cosmology}: the exact-numerical representation
of the large difference in cosmological redshift between quasars
and galaxies when physically connected according to gamma spectroscopic
evidence
(as due to Santilli's isodoppler shift within the huge quasars
chromospheres
according to which light exits quasars already
redshifted to the value of the associated galaxy)
[10j]; the first and still the only available numerical representation
of the
internal quasars redshift and blueshift [10k]; the elimination of
the missing mass in the universe [5i]; and other verifications.

{\bf D) Superconductivity}: the first and only known model of the Cooper
pair
with an {\it explicitly attractive force} between the {\it
two identical electrons} of the pair in remarkable agreement with
experimental
data [10l,10m]]; and other verifications.

{\bf E) Chemistry}: the first known representation of the binding
energy,
electric and magnetic moments, and other characteristics
of the hydrogen, water and other molecules which are {\it exact to the
seventh digit} (quantum chemistry still
misses about 2\% of the binding energies,
with much bigger insufficiencies in electric and
magnetic moments, which at times even have the wrong sign) [10n,10o]];
several independent experimental verifications of the
prediction of a {\it new chemical species} composed of conventional
molecules
and atoms under a new magnetic bond originating
from the polarization the orbits of the valence
electrons (which produce a field about 1,400 times stronger than
nuclear magnetic fields) and related new industry
of magnetically polarized gases
[10p,10q]; and other verifications.
\vskip 0.50 cm

\noindent {\large \bf 4. Direct Universality of the L-P-S Isosymmetry}

\noindent The Lorentz-Poincar\'e-Santilli (L-P-S) isosymmetry
is directly universal for closed-isolated systems
verifying conventional total conservation laws, with linear and
nonlinear, local and nonlocal and potential-Hamiltonian as well as
nonpotential-nonhamiltonian internal dynamics, where: 1) the
verification
of conventional total conservation laws is established by the fact that
the
generators of the isopoincar\'e symmetry are conventional (see next
section); 2)
all linear, local and potential forces are represented via the
conventional Hamiltonian; and 3) all "non-non-non" effects are
represented with the isounit.

The understanding of isotopic theories requires at least a rudimentary
knowl;edge of
the above direct universality, if nothing else, to prevent the
alternative use for the same problem of theories with
catastrophic physical inconsistencies. The best way to achieve
a rapid and intuitive understanding is the geometric way. In turns this
is useful to understand the local isomorphism of the conventional and
isotopic spacetime symmetries even prior to their treatment in the next
section.

As it is well known, the Minkowskian geometry and the
 rotational-Lorentz-Poincar\'e symmetry can only
characterize perfectly spherical and perfectly rigid
shapes $r^2 = x^2 + y^2 + z^2$ which are geometrically
represented via the {\it unit of the Euclidean subspace}
$I = Diag. (1, 1, 1)$.
In fact, any shape other than the perfect spere and any deviation from
its perfect rigidity imply the collapse of the pillar of
spacetime symmetries, the rotational symmetry.

Santilli [5b] achieves the most general possible,
signature preserving
(compact) deformation of the sphere
while preserving the rotational symmetry as {\it exact}.
Recall that the Euclidean unit represents in a dimensionless form
the basic units of length along the three space axes,
$I = Diag. (1 cm^2, 1 cm^2, 1 cm^2)$, where the square is evidently
due to quadratic character of the interval. Then, jointly with
the lifting of the sphere into the most
general possible spheroidal ellipsoids, Santilli lifts
the corresponding units by the {\it inverse} amount,
$$
r^2 = x^2 + y^2 + z^2\rightarrow
r^{\hat 2} = x^2/n_1^2 + y^2/n_2^2 + z^2/n_3^2,
$$
$$
I = Diag. (1 cm^2, 1 cm^2, 1 cm^2)\rightarrow
\hat I = Diag. (n_1^2 cm^2, n_2^2 cm^2, n_3^2 cm^2),
\eqno (4.1)
$$

It is then easy to see that the deformed sphere is indeed
the perfect sphere in {\it isoeuclidean space}
 $\hat E(\hat r,\hat {\delta},\hat R), \hat r = r\times \hat I, \hat
{\delta}
= Diag (n_1^{-2}, n_2^{-2}, n_3^{-2})\times \delta$, called
the {\it isosphere} [5]. In fact, each semiaxis
is subjected to the lifting
$1_k\rightarrow n_k^{-2}$; but the corresponding units
are lifted by the inverse amount, $1_k cm^2\rightarrow n_k^2 cm^2$.
This implies  the
preservation of the {\it original numerical value} of the semiaxes in
isospace.
The latter occurrence is due to the fact
that all invariants are elements of the
underlying field. As such, they should be written in general $r^2 =
(x^2 + y^2 + z^2)\times I = n\times I$, where I is
the unit of the field. The preservation of the perfect spheridicity
under the liftings (4.1) then follows, as established by
invariant (3.3). The extension to shapes other than
spheroidal ellipsoids is easily achieved via
{\it nondiagonal positive-definite isounits} (see
monograph [5p] for brevity).

The understanding of the perfect spheridicity
of $r^{\hat 2} = x^2/n_1^2 + y^2/n_2^2 + z^2/n_3^2$
in isospace then permits the understanding of the property that,
contrary
to all popular beliefs throughout this century,
{\it the rotational symmetry remains indeed perfectly exact for all
infinitely possible compact deformations of the sphere. }

By comparison, the representation of extended particles
by Y. S. Kim [4a] is a particular case of Santilli
broader representation [5a,5b].
As indicated earlier, the former can only occur for perfectly
spherical and perfectly rigid shapes, while the latter occurs for
arbitrarily nonspherical and deformable shapes. Whenever
the former is extended to include the latter, the catastrophic physical
inconsistencies
of Sect. 2 are activated, trivially, because the map from a
perfectly spherical to a nonspherical shape is necessarily noncanonical-
nonunitary.

The restriction of particle/ charge distributions
to be perfectly spherical and perfectly rigid has
rather serious physical implications. As an illustration, it
prohibits the achievement (indicated in Sect. 3) of an exact
representation of
nuclear magnetic moments (which require precisely a nonspherical
deformation
of nucleons), and other applications.

This illustrates the comment of Sect. 2 to the effect that the work of
Ref. [4a] and literature quoted
therein is invaluable to establish the maximal capability of
the conventional realization of the Lorentz-Poincar\'e axioms,
with the clear understanding necessary not to exit science that,
by no means, it is the final theory. At any rate, the little groups
of Ref. [4a] are contained as a particular case of Refs. [5];
Ref. [4a] departs from the Lorentz-Poincar\'e teaching of "invariance"
in favor of a "covariance, while Refs. [5] restore the "invariance"
in its entirety; and, finally, the entire mathematical treatment of
Ref. [4a] can be used for the representation of the missing
nonspherical and deformable shapes via Santilli's re-interpretation of
all symbols as being of isotopic character.

The representation via the Lorentz-Poincar\'e-Santilli isosymmetry
of extended, nonspherical and
deformable shapes is only the beginning of its direct universality.
The next important applications are
the representation of arbitrary speeds of light
while preserving on isospace $\hat M$ of the maximal causal speed of M
(the speed of light in vacuum),
and consequential preservation of the light cone.
Contrary to a popular belief throughout this century, this
feature establishes that {\it the Lorentz-Poincar\'e symmetry is exact
for arbitrary speeds of light}.

Recall that Minkowski originally
wrote his metric in the form $\eta = Diag. (1, 1, 1, -c_o^2)$.
Therefore, the fourth component of the Minkowski metric represents in a
dimensionless form the unit $cm^2/sec^2$, and the metric explicitly
reads
$\eta = Diag. (1, 1, 1, - 1 cm^2/sec^2)$.
In the isominkowskian space, Santilli [5a] considers: 1) the lifting
from $c_o^2$ to an
arbitrary local speed $c^2 = c_o^2/n_4^2(x, v, d, \tau,\psi, ...)$,
where
n is the local index of refraction; and 2) the joint lifting the
unit by the {\it inverse} amount. It is then evident that the dual
lifting
$$
\eta = Diag. (1, 1, 1, -c_o^2)\rightarrow
\hat \eta = Diag. (1, 1, 1, - c_o^2/n_4^2(x, v, d, \tau, \psi, ...)),
$$
$$
I = Diag. (1, 1, 1, 1 cm^2/sec^2)\rightarrow
\hat I = Diag. (1, 1, 1, -n_4^2 cm^2/sec^2),
\eqno (4.2)
$$
implies the preservation of the maximal causal speed $c_o$ on isospaces
over
isofields (that is, when considered with respect to $\hat I$).

It is evident that,
when projected in the ordinary spacetime (that is, when considered with
respect to I) the isorelativistic  theory represents
the local speed $c = c_o/n_4$.

The additional use of the isosphere then yields {\it Santilli's light
isocone}
[5] which is the perfect cone in isospace. The abstract identity
pf the isocone with the
conventional cone is such that even the
characteristic angles of the two cones coincide (to prevent insidious
misrepresentation, one should know that
the proof of this
occurrence requires the use of the isotrigonometric and
isohyperbolic functions [5p], the use of conventional
mathematics in isospace being as fundamental inconsistent as the
treatment
of conventional theories via isomathematics ).

Recall that speeds $c < c_o$ are known
since the discovery of the refraction of light, while
speeds $c > c_o$ have been
experimentally measured in recent times, and can be
considered as established for all interior
media of high density,
such as those in the interior of hadrons or of stars [11].

It then follows that
{\it the isopoincar\'e symmetry extends the
applicability of the conventional Einsteinian
axioms, from their sole validity for speeds of light in vacuum,
 to arbitrary speeds within physical media.} To put it differently,
the special relativity becomes "directly universal" when
formulated in the form today known as
{\it Santilli's isospecial relativity} [5-10].

A glimpse at the applications may be of some value here. Nowadays the
light
cone is used for all calculations outside gravitational horizons and
in similar conditions. However, in the outside of gravitational horizon
we have something
dramatically different than the vacuum. In fact we have a region
of space filled up of hyperdense chromospheres in which the speed of
light, first of all, positively is not that in vacuum and, second
it is locally variable. As a result, outside
gravitational horizons we have highly deformed "cones". Santilli's
light isocone permits a more scientific study of these regions thanks
precisely to the admission of local arbitrary speeds.

Note that the traditional hopes of representing light within physical
media via
photons scattering among molecules (to maintain the speed of light
in vacuum $c_o$)
has been discredited by the recent
experimental evidence of speeds {\it bigger} than $c_o$. At any rate,
we are referring here to a purely classical event (the propagation of
electromagnetic waves within physical media at speeds $c < c_o$) which,
as such,
cannot be credibly reduced to {\it photons in second quantization}
without a prior {\it classical} representation.

By no means the above topics are pure semantic,
because they have deep implications in the {\it numerical values}
of physical characteristics throughout the universe.

As an illustration in the macrocosm, the belief of the validity of the
conventional light cone
everywhere in the universe leads to the current beliefs of the size
of the universe (generally derived from the cosmological redshift).
However, the admission of the physical reality that the speed of light
decreases within astrophysical chromospheres implies the necessary
consequence that light exits said chromospheres {\it already redshifted}
(see Santilli's companion paper [10b] for the explicit treatment).
The decrease of the currently
believed size of the universe is then simply incontrovertible.

As an illustration in the microcosm, the possibility to stimulate the
decay
of the neutron and related new forms of energy mentioned
in Sect. 3, originates
precisely from the admission that light travels in the hyperdense '
medium inside hadrons at a speed different than that in vacuum.

Next, it is easy to see that all infinitely possible {\it Riemannian}
metric g(x) are simple {\it particular cases} of the
isometric $\hat \eta(x, v, d, \tau, \phi, ...)$. In fact,
Santilli [5j] has introduced
the novel {\it isominkowskian formulation of
gravitation and general relativity}
based on the {\it Minkowskian factorization of the Riemannian metric}
$$
g(x) = \hat T_{grav.}(x)\times \eta, \hat I_{grav.}(x) = 1/\hat
T_{grav.},
\eqno(4.3)
$$
and consequential reconstruction of the entire
Riemannian formalism into such a form to admit $\hat I_{grav.}$ as the
correct left and right new unit.

This result was possible thanks to the construction
of the novel
{\it isominkowskian geometry} [loc. cit.]
as a symbiotic unification of the Minkowskian
and the Riemannian geometries indicated in Sect. 3.

A visible illustration is the
{\it isominkowskian formulation of Schwarzschild} [5j]
$$
\hat d s^{\hat 2} = \hat d \hat r^2 + \hat r^{\hat 2}\hat \times
 (\hat d {\theta}^2 + \hat {sin}^2\theta
\times \hat d{\phi}^2) -
\hat d\hat t^{\hat 2}\times c_o^2,
$$
$$
\hat d r = \hat I_s\times dr, \hat dt = \hat I_t\times dt,
\hat I_s = (1 - 2M/r)^{-1}, \hat I_t = 1 - 2M/r,
\eqno(4.4)
$$
where, as one can see, curvature disappears completely because it
is embedded in the differential calculus, thus permitting the
regaining of a fully {\it minkowskian} structure for
{\it Schowarzschild's metric.}

Santilli's isominkowskian formulation
of gravity implies considerable structural novelties, such as:

1) The formulation, for the first time to our knowledge,
of gravitation under the rigid validity of a {\it symmetry}
(rather than the usual covariance), which results to be
isomorphic to the Poincar\'e symmetry.

2) The abandonment of the conventional curvature in favor of {\it
isoflatness},
that is, flatness in isospace, as transparent in the reformulation
(4.4).

3) The unification of the special and general relativities
which are now differentiated by the {\it unit}, rather than by the
geometry,
while the underlying geometry remains unchanged. Equivalently, we can
say that
Santilli's isominkowskian representation of gravity extends the
direct universality
of the {\it special} relativity to include {\it gravitation} where
nobody looked
before, in the unit of the theory.

As shown in Ref. [5j], this reformulation of gravity permits the
resolution
of at least some of the controversies in gravitation that have raged
through
this century, such as:

A) The reconstruction on isospaces over isofields of
full canonicity or unitarity (isocanonical or
isounitary laws). In turn,
this permits the regaining for gravitation of invariant
basic units of measurements,
the preservation of Hermiticity-observability at all times, and resolves
the
other physical inconsistencies
of general relativity indicated in Sec. 2.

        B) The compatibility between relativistic and gravitational
total conservation laws, which is established via the mere
visual inspection that the generators
of the isopoincar\'e and conventional Poincar\'e
symmetries coincide (see next section).
It is instructive to compare this geometric-
algebraic simplicity with the complexity
of the conventional proof of total conservation laws in general
relativity.

C) The existence, for the first time to our knowledge, of a
consistent {\it relativistic} limit
of Riemann, which is now established via the limit
$\hat I_{grav.}\rightarrow I$);
 and other resolutions.

Moreover, the regaining of flat gravity in isospace permitted
the achievement of
the first grand unification of electroweak and
gravitational interactions which is axiomatically consistent
[5g,5h]. In Ref. [5h], p. 324,
one can read the viewpoint according to which:
"gravitation has always been present in unified gauge theories. It did
creep
in un-notices because occurring where
nobody looked for, in the 'unit' of gauge theories".
Electroweak gauge theories can be identically formulated
on isospaces. Then, gravity is contained
in the unit of the isosymmetries $\hat U(2)\times \hat U(1)$.

The resulting {\it iso-grand-unification} (IGU) also identifies the
technical reasons for the axiomatic inconsistency of other unified
theories.
In fact, electromagnetic interactions, as well as electroweak
interactions
in general, rigorously follow the Poincar\'e symmetry. Any attempt at
adding gravitation without a {\it symmetry} is then doomed to failure.
Via his reformulation of gravitation in such a way to admit
a symmetry isomorphic to
Poincar\'e, Santilli has resolved the apparently deepest
historical obstacle against
a grand unification. This illustrates a reason why all
attempts initiated by Einstein and
continued by many scholars were doomed to fail from their
very foundations. Other resolutions of structural incompatibilities
between gauge and gravitational theories are related to
the treatment of antimatter and will di discussed later on.

Santilli has also pushed his studies to the formulation
of the novel
{\it isocosmology} [5i] which brings the validity of the studies by
 Lorentz, Poincar\'e, Einstein, Minkowski,
and others to a true "universal" level, that
of cosmological character inclusive of gravitation.
Some of the rather intriguing implications of the isocosmology are:
the elimination of the need for a "missing mass" in the universe
because the energy equivalence is now $E = m\times c^2 = c_o^2/n_4^2$,
rather than $E = m\times c_o^2$, with an average value of c for
galaxies, quasars and the universe
in general much bigger than $c_o$ when considering all {\it interior}
gravitational problems; a significant reduction of the
currently believed dimension of
the universe (indicated earlier in this section); and other intriguing
features.

By no means this exhausts all the applications of the isominkowskian
geometry, isopoincar\'e symmetry and isospecial relativity. The next
application is the study of relativistic and
gravitational {\it interior} problems at large,
e.g., the formulation of the {\it Schwarzschild solution for
interior problems
with local speed} $c = c_o/n_4(x, v, d, \tau, ...)$ [5j].

In particular, {\it gravitational horizons (singularities) result to be
the zeros of the time (space) component of the isounit},
as one can verity from structures (4.4). This
is not a mere mathematical curiosity. Gravitational collapse
is one of the most complex physical events in the universe, with
the consequentially most complex possible dependence of the
metric on all conceivable local
quantities. In particular, as typical of interior
trajectories (such as those of missiles in atmosphere),
we must expect in the gravitational collapse
 an {\it arbitrary dependence of the metric
in the velocities} (which is simply impossible for Riemann), nonlocal-
integral effects due to total wave-overlappings of a large
number of wavepackets in a small region
of space (which effects are precluded by the topology
of Riemann), and
interactions which violate the integrability conditions
for the existence of a Lagrangian (the
{\it conditions of variational selfadjointness} [6g])
which is also beyond any dream of representation via Riemann.
In short, the assumption of the Riemannian
geometry as being exact for gravitational collapse in general
and for the study of gravitational singularities in particular,
is so questionable to imply exiting science.

All the above interior features are directly represented by
Santilli's isominkowskian
geometry, thus permitting, for the first
time, more realistic studies of interior gravitational
problems in general, gravitational collapse in particular, and related
topic, such as whether or not the universe started from a "big bang".

The direct universality of the isorelativistic formalism is also
established by the fact that it admits
as a particular case all infinitely possible {\it Galilean}-types
of
space and time. They are evidently admitted under the particular
 Kronecker structure of the isounit
$$
\hat I = \{\hat I_{Space}\}\times \hat I_{time}.
\eqno (4.5)
$$
with consequential factorization of the isopoincar\'e
symmetry into the {\it  isogalilean symmetry} [5m.5n].The latter
aspects
are not considered here for brevity.
\vskip 0.50 cm

\noindent {\large \bf 5. Explicit form of trhe L-P-S Isotransforms.}

\noindent In this section we outline the operator version of the
L-P-S
isosymmetry $\hat P(3.1)$, with particular reference to the explicit
form of
the isosymmetry transformations (called {\it isotransforms}).

In inspecting this section the reader should keep in kind
that it provides a {\it direct operator theory of gravity}
called {\it operator isogravity} [5j,5l] under the sole restriction of
the
isominkowskian metric $\eta(x, v, d, ]\tau, \psi, ...)$
to be the conventional Riemannian metric g(x). The resulting
new theory coincides at the abstract level with the conventional
relativistic quantum mechanics, thus preserving {\it all} its
properties.
This occurrence is sufficient, alone, to establish the axiomatic
consistency of
operator isogravity beyond scientific doubt. Such a
'consistency should then be compared,
for scientific objectivity, with the catastrophic physical
inconsistencies of quantum gravity outlined in Sect. 2.

The clear understanding is that the operator isorelativistic theory
outlined
below has applications much beyond that of the mere operator formulation
of gravity (Sect. 4).

The isosymmetry $\hat P(3.1)$ is  characterized  by the
{\it conventional} ten generators and parameters of P(3.1), only lifted
into
their corresponding forms on isospaces over
isofields, {\it plus} the 11-th generator $\cal S$ for the
new symmetries (3.3), (3.4) with parameter $\rho$
$$
X   = \{X_{k}\}  = \{M_{\mu\nu}   =
x_{\mu}\times p_{\nu} - x_{\nu}\times p_{\mu}, p_{\alpha}, \cal S\}
\rightarrow \hat {X} = \{\hat {M}_{\mu\nu} = \hat {x}_
{\mu}\hat {\times}\hat {p}_{\nu} -
\hat {x}_{\nu}\hat {\times}\hat {p}_{\mu}, \hat {p}_{\alpha}, \hat {\cal
S}\},
$$
$$
 w = \{w_{k}\} = \{(\theta ,v),a, \rho \}
 \in R \rightarrow \hat{w}
= w\times \hat{I} \in \hat{R}(\hat{n},\hat +,\hat{\times }),
\mu ,\nu = 1,2,3,4; k = 1,2,...,11.
\eqno(5.1)
$$

The isotopies
preserve the  original  connectivity properties of the Lorentz group
L(3.1)
[5-8].  The $\hat P(3.1)$    isosymmetry is  then   given   by
$$
\hat{P}(3.1)   =
[ \hat{L}(3.1)\hat{\times }\hat{T}(3.1)]\times \hat {\cal S}
\eqno (5.2)
$$,
where $\hat{L}(3.1)$ is  the
{\it  isolorentz group} [5a],  $\hat{T}(3.1)$ is  the
group   of {\it   isotranslations} [5e], and
$\hat {\cal S}$ is the one-dimensional group
of symmetries (3.3), (3.4), which is evidently in the center
of the isogroup. Note that
the latter essentially acts as the isotopies of
the conventional "scalar extensions" of Lie's symmetries,
as familiar for the Galilei's (but not for the Poincar\'e) symmetry.
Santilli has identified their origin in the change of
the unit of the underlying field and found its explicit symmetry
transforms.

The {\it  isoexponentiation} characterized  by   the
 {\it   Poincar\'{e}--Birkhoff-Witt-Santilli
\- theorem}\-   [6a,6g,7g,8] of  the    underlying enveloping
isoassociative
algebra  $\hat {\cal A}(\hat P(3.1))$
$$
\hat{e}^{A} = \hat{I}  +
A/1! + A\hat{\times }A/2! + . . . = (e^{A\times\hat{T}})\times\hat{I}
\eqno (5.3)
$$
permits to write the connected component  of the L-P-S isosymmetry
$\hat{P}_{o}(3.1) = \hat {SO}(3.1)\hat {\times} \hat T(3.1)$  in
the form
$$
\hat{P}_{o}(3.1):\hat{A}(\hat{w})  =
\Pi_{k=1,...,10}\hat{e}^{i\times X\times w}               =
(\Pi_{k}e^{i\times X \times\hat{T}\times w}) \times \hat{I} =
\tilde{A}(x, v, d, \tau, \psi, ...)\times \hat{I}.
\eqno (5.4)
$$

Note the  appearance of the  isotopic element
$\hat{T}(x, v, d, \tau, \psi, ...))$
in the {\it exponent} of   the group structure. This illustrates   the
nontriviality   of the Lie-Santilli isotheory and,
in particular, its   {\it nonlinear, nonlocal and nonunitary} characters
in its projection on conventional spaces over conventional fields.
Intriguingly,
 the   isopoincar\'{e} symmetry recovers linearity,
 locality and unitarity  on
$\hat{M}$ over $\hat{R}$.

Conventional linear transforms on   $M$ {\it violate} isolinearity  on
$\hat{M}$  and  must then be    replaced with the {\it  isotransforms}
$$
\hat{x}' = \hat{A}(\hat{w})\hat{\times }\hat{x} =
\hat{A}(\hat{w}. \hat x, \hat v, ...)\times\hat{T}(x)\times\hat{x} =
 \tilde{A}(w, x, v, ...)\times\hat{x}.
\eqno (5.5)
$$

The  preservation of   the
original     ten dimensions   is   ensured     by     the  {\it
isotopic
Baker--Campbell--Hausdorff    Theorem} [6a].  Structure  (5.4) then
forms a connected {\it Lie--Santilli isogroup} with laws
$$
\hat{A}(\hat{w})\hat{\times}\hat{A}(\hat{w}') =
\hat{A}(\hat{w}')\hat{\times}\hat{A}(\hat{w}) = \hat{A}(\hat{w} +
\hat{w}'),\hat{A}(\hat{w})\hat{\times}\hat{A}(-\hat{w}) = \hat{A}(0) =
\hat{I} = \hat {T}^{-1}.
\eqno(5.6)
$$

The  use    isodifferential calculus on   $\hat{M}$   then  yields the
L-P-S isoalgebra $\hat{\bf p(3.1)}$ [5]
$$
\hat{M}_{\mu\nu     },\hat{}\hat{M}_{\alpha\beta  }]   =   i  \times
(\hat{\eta }_{\nu\alpha } \times \hat{M}_{\mu\beta } -
\hat{\eta }_{\mu\alpha } \times \hat{M}_{\nu\beta } -
\hat{\eta }_{\nu\beta } \times \hat{M}_{\mu\alpha } +
\hat{\eta }_{\mu\beta } \times \hat{M}_{\alpha\nu }),
$$
$$
\hat{M}_{\mu\nu }, \;\;   \hat{p}_{\alpha   }]  = i   \times
(\hat{\eta
}_{\mu\alpha } \times \hat{p}_{\nu } -
\hat{\eta }_{\nu\alpha} \times \hat{p}_{\mu}),
$$
$$
[\hat{p}_{\alpha },\hat{}\hat{p}_{\beta }] =
[\hat M_{\mu\nu}\hat {,} \hat {\cal S}] =
[\hat p_{\mu}\hat {,} \hat {\cal S} ] = 0
$$
$$
\hat {p}_k\hat {\times} |\hat {\psi}> = i \hat {\partial}_k |\hat
{\psi}>,
[A\hat {,}B]   =     A\times\hat{T}\times     B     -
B\times\hat{T}\times  A
\eqno (5.7)
$$
where   $[A\hat {,}B]$   is   the  {\it   Lie-Santilli isoproduct}
(first
proposed in [6b]),  which does  indeed satisfy  the Lie axioms   in
isospace, as one can  verify.

Note for the particular case $\hat {\eta} = g(x)$ the
appearance of the {\it  Riemannian
metric as the 'structure functions'}.  Note also  that the
{\it momentum components
isocommute} (while they are  notoriously non--commutative for
quantum gravity). This
confirms the achievement of the isoflat representation of gravity
indicated in Sect. 4, which is seemingly mandatory
to achieve a consistent
grand unification of gravity with other interactions.

The local isomorphism ${\bf \hat{p}(3.1) \approx p(3.1)}$ is ensured
by the  positive--definiteness  of $\hat{T}$. In  fact,  the use of the
generators    in  the  form $\hat{M}^{\mu   }_{\nu    } = \hat{x}^{\mu
}\hat{\times }\hat p_{\nu} -
\hat{x}^{\nu}\hat{\times}\hat{p}_{\mu }$ yields the {\it conventional}
structure constants under a {\it generalized} Lie  product, as one can
verify.  The  above   local isomorphism is   sufficient,  per se',  to
guarantee the axiomatic consistency of the L-P-S isosymmetry
and all applications in which it is exact, including operator
isogravity.

The {\it isocasimir invariants}  of ${\bf \hat{p}(3.1)}$ are  the simple
isotopic    images  of  the  conventional    ones
$$
C^{o}  =  \hat{I} =
[\hat{T}(x, v, d, \tau, \psi, ...)]^{-1},
$$
$$
 C^{(2)} = \hat{p}^{\hat{2}} =
\hat{p}_{\mu }\hat{\times }\hat{p}^{\mu } =
\hat{\eta}^{\mu\nu} \times \hat{p}_{\mu}\hat{\times}\hat{p}_{\nu},
$$
$$
C^{(4)} =
\hat{W}_{\mu }\hat{\times }\hat{W}^{\mu }, \hat{W}_{\mu } =
\in _{\mu\alpha\beta\pi }\hat{M}^{\alpha\beta }\hat{\times}\hat{p}^{\pi
}
\eqno (5.8)
$$.
>From     them, one can   construct     any needed {\it
isorelativistic equation}, such as the {\it Dirac-Santilli isoequation}
[5f]
$$
(\hat{\gamma     }^{\mu     }\hat{\times    }\hat{p}_{\mu     }    +
\hat {i} \hat {\times}\hat{m})\hat{\times
}|>  =   [\hat{\eta    }_{\mu\nu   }(x, v, ...)  \times
{\hat{\gamma }}^{\mu} \times \hat{T}
\times\hat{p}^{\nu} - i\times m \times \hat{I}]
\times \hat{T} \times |> = 0,
$$
$$
\{\hat{\gamma }^{\mu },\hat{}\hat{\gamma }^{\nu}\} =
\hat{\gamma }^{\mu}\times\hat{T}\times\hat{\gamma }^{\nu} +
\hat{\gamma }^{\nu}\times\hat{T}\times\hat{\gamma }^{\mu } =
2\times \hat{\eta}^{\mu\nu },
\hat{\gamma }^{\mu} = \hat{T}_{\mu\mu }^{1/2}\times
\gamma^{\mu}\times\hat{I}\;
({\rm no}\; {\rm sum}),
\eqno (5.9)
$$
where $\gamma^{\mu  }$  are the  conventional  gammas and $\hat{\gamma
}^{\mu  }$ are    the  {\it isogamma   matrices}.

Note  that, again for the particular case $\eta(x, v, d, ...) = g(x)$,
 {\it the
anti-isocommutators  of   the    isogamma matrices  yield   twice  the
Riemannian    metric},  thus   confirming  the  representation of
gravitation in   the {\it structure} of  Dirac's
equation.

As an illustration, we   have the {\it Dirac--Schwarzschild
isoequation}   characterized by    $\hat{\gamma  }_{k}    =
(1-2M/r)^{-1/2}\times\gamma_{k}\times\hat{I}$ and  $\hat{\gamma }_{4}
=   (1-2M/r)^{1/2}\times\gamma^{4}\times\hat{I}$.   Similarly  one can
construct  the   isogravitational version of   all other  equations of
relativistic quantum mechanics.

These equations are not a mere mathematical curiosity because they
establish the compatibility of operator isogravity
with  experimental data
 in view   of the much  smaller value  of gravitational  over
electromagnetic, weak and strong interactions.
The isotopic unification of the
special and   general   relativities is,  therefore,   compatible with
experimental evidence at both classical and operator levels.

The explicit form of the isosymmetry transformations are
by:

{\bf 1) Isorotations} [5b],
which can be computed from isoexponentiations  (5.4)
resulting in the explicit form in
the  (x,y)--plane (were we  ignore hereon the factorization of $\hat{I}$
for simplicity)
$$
x' =
x\times\cos(\hat{T}_{11}^{\frac{1}{2}}\times\hat{T}_{22}^{\frac{1}{2}}
\times\theta_{3}) -
y\times\hat{T}_{11}^{-\frac{1}{2}}\times\hat{T}_{22}^{\frac{1}{2}}
\times
\sin(\hat{T}_{11}^{\frac{1}{2}}\times\hat{T}_{22}^{\frac{1}{2}}
\times\theta_{3}),
$$
$$
y' =
x\times\hat{T}_{11}^{\frac{1}{2}}\times\hat{T}_{22}^{-\frac{1}{2}}\times
\sin(\hat{T}_{11}^{\frac{1}{2}}\times\hat{T}_{22}^{\frac{1}{2}}\times
\theta_{3}) + y\times\cos(\hat{T}_{11}^{\frac{1}{2}}\times\hat{T}_{22}^
{\frac{1}{2}}\times\theta_{3}),
\eqno (5.10)
$$
(see [5p]   for     general  isorotations    in    all there   Euler
angles).   Isorotations  (5.10)   leave   invariant  all   ellipsoidical
deformations of the sphere indicated in Sect. 4, as the reader
is encouraged to verify. The local isomorphism between $\hat O(3)$ and
O(3)
then confirms the perfect spheridicity of ellipsoids on isospace
(the isosphere).

Note that the space components of all gravitational theories
characterize an isosphere when reformulated on
isoeuclidean spaces over isofields.

{\bf 2) Isolorentz transformations} [5a],
which are characterized by the
isorotations and the {\it isoboosts}, e.g.,
in the $(3,4)$--plane
$$
x^{3}{}' = x^{3}\times\sinh(\hat{T}_{33}^{\frac{1}{2}}\times
\hat{T}_{44}^{\frac{1}{2}}
\times v) - x^{4}\times\hat{T}_{33}^{-\frac{1}{2}}\times\hat{T}_{44}^
{\frac{1}{2}}\times
\cosh(\hat{T}_{33}^{\frac{1}{2}}\times\hat{T}_{44}\times       v)     =
$$
$$
 \tilde{\gamma}\times
(x^{3}-\hat{T}_{33}^{-\frac{1}{2}}\times\hat{T}_{44}^{\frac{1}{2}}\times
\hat{\beta}\times  x^{4})
$$
$$
x^{4}{}' = -x^{3}\times\hat{T}_{33}\times
c_{0}^{-1}\times\hat{T}_{44}^{-\frac{1}{2}}\times\sinh
(\hat{T}_{33}^{\frac{1}{2}}\times\hat{T}_{44}\times v)  +  x^{4}\times
\cosh (\hat{T}_{33}^{\frac{1}{2}}\times\hat{T}_{44}^{\frac{1}{2}}\times
v) =
$$
$$
\tilde{\gamma }\times (x^{4}-\hat{T}_{33}^{\frac{1}{2}}
\times\hat{T}_{44}^{-\frac{1}{2}}\times\tilde{\beta}\times x^{3}),
$$
$$
\tilde{\beta} = v_{k}\times\hat{T}_{44}^
{\frac{1}{2}}/c_{0}\times\hat{T}_{44}^{\frac{1}{2}},
\tilde{\gamma} = (1-\tilde{\beta}^{2})^{-\frac{1}{2}}
\eqno (5.11)
$$

Note that the above isotransforms are formally similar to
the Lorentz transforms,
as expected from their isotopic character.

Note also that all (3+1)-dimensional Riemannian
models of gravity, when subjected to
their isominkowskian reformulation, characterize precisely light
isocones
on $\hat M$ over $\hat R$.  All possible gravitational models
are therefore unified into one single primitive geometric notion.

{\bf 3) Isotranslations} [5e], which
can be written
$$
x' = (\hat{e}^{i\times\hat{p}\times a})\hat{\times}\hat{x} =
[x + a\times A(x, v, d, ...)]\times\hat{I},
\hat{p}' = (\hat{e}^{i\times\hat{p}\times a})\hat{\times}\hat{p} =
\hat{p},
$$
$$
A_{\mu} = \hat{T}_{\mu\mu }^{1/2} + a^{\alpha }\times[\hat{T}_
{\mu\mu }^{1/2},\hat{}\hat{p}_{\alpha }]/1! + ....
\eqno (5.12)
$$
and they are also nonlinear, as expected.

{\bf 4) Isoinversions} [5e], which   are    given      by
$$
\hat{\pi}\hat{\times}x =
\pi\times x = (-r,x^{4}), \hat{\tau }\hat{\times }x = \tau\times x =
(r,-x^{4})
\eqno (5.13)
$$
where $\hat{\pi } = \pi\times\hat{I},  \hat{\tau } = \tau\times\hat{I}$,
and  $\pi$, $\tau$ are the   conventional inversion operators.

Despite
such a simplicity, the physical implications of the
isoinversions are nontrivial because of
{\it the possibility    of  reconstructing as  exact
discrete  symmetries when believed  to be  broken}, which
can be achieved by
embedding all symmetry breaking
terms in the isounit [5p].

One should be aware that the reconstruction
of exact spacetime and internal symmetries is a rather
general property of the Lie--Santilli  isotheory, thus holding also for
continuous  symmetries. In fact, contrary  to popular beliefs, this
section shows
 that  {\it the Lorentz  and Poincar\'{e} symmetries  are exact
for gravitation}.

{\bf 5) Isoselfscalar transforms} [5h], which
are characterized by invariances (3.3)-(3.4), i.e.,
$$
\hat{I} \rightarrow \hat{I}' = \rho^{2}\times\hat{I},
\hat{\eta} \rightarrow \hat{\eta}' = \rho^{-2}\times\hat{\eta},
\eqno (5.14)
$$
where $\rho$ is the parameter characterizing
the novel 11-th dimension.

The implications of the 11-th invariance of spacetime is now clear: it
permits the achievement of a consistent grand unification of
gravitation and electroweak interactions according to a mechanism
essentially
equivalent to the unification of electromagnetic and weak
interactions, the generalization of the parameter $\rho$
into the positive-definite function $\hat T_{grav.}(x)$ and
the rule
$$
(x^\mu\times \eta_{\mu\nu}\times x^\nu )\times I =
\{x^\mu\times [\hat T_{grav.}(x)_{\mu}^
{\rho}\times\eta_{\rho\nu}]\times x^\nu \}\times
[\hat T_{grav.}^{-1}(x)\times I] =
$$
$$
[x^\mu\times g_{grav.\mu\nu}(x)\times x^\nu ]\times \hat
I_{grav.}(x).
\eqno (5.15)
$$
where the equality holds for the two sides computed
in their respective spaces and fields.

By looking in retrospect, we can say that the
apparent reason why a grand unification was not achieved
during this century until recently was theabsencek of one
dimension in the basic symmetry of spacetime [5g,5h].

The above results can be summarized with the following:
\vskip 0.50 cm

{\it THEOREM 3 (Direct
universality of the L-S-P isosymmetry [5]):
The 11-dimensional,
Lorentz-Poincar\'e-Santilli isosymmetry on
isominkowski spaces
over real isofields with common, $4\times 4$-dimensional,
positive-definite isounits constitutes the
largest possible isolinear, isolocal and isocanonical-isounitary
invariance
of isoseparation (3.1), thus being directly universal for all possible,
Galilean, relativistic or gravitational, interior and exterior
spacetime theories with a well behaved symmetric metric.}
\vskip 0.50 cm

It should be stressed that for any
arbitrarily given diagonal metric $\hat {\eta} = \hat T\times \eta$
 {\it there is nothing to compute} because one merely {\it plots}
the $\hat{T}_{\mu\mu}$ terms in the above given isotransforms.
The invariance of interval (3.1)
is then assured by Theorem 3. The $(2+2)$--de Sitter or other diagonal
cases can be derived from the above
theorem via mere changes of signature or dimension of the isounit.
The case of nondiagonal metrics will be considered in the next section.

\vskip 0.5 cm

\noindent {\large \bf 6. Santilli's geno- and
hyper-symmetries and their isoduals.}

\noindent The limitations of fundamental theories can be
best identified in the writings of their originators
(rather than of their followers). For instance, Lorentz was the first
to identify the insufficiency of his historical symmetry for the
speeds of light of the physical reality and conducted the first
search for a possible broader symmetry for speeds $c < c_o$ [11a] (while
his followers proclaimed for the rest of this century the "universal
constancy of the speed of light").

Receptive to the and other historical teachings, Santilli
has identified the major limitation of
his isotopies as bneing {\it their inability
to permit axiomatically correct studies on irreversibility }.
In fact, being "axiom-preserving", the isotopies preserve the original
inability by the L-P axioms to describe irreversibility.

For these reasons, Santilli conducted his studies via the
broader {\it genomathematics} initiated in his Ph.D. thesis back
in 1967, [12a] and then studied in various works [5a,5b,5c, 5l,6e,12].
Invariance was achieved for the first time
in memoir [12e] of 1997. This is
a broader mathematics possessing a {\it Lie-admissible} (rather than
Lie-isotopic)
structure (a generally nonassociative
 algebra U with abstract product ab is said to be Lie-
admissible when the attached algebra $U^-$, which is the same
vector space as U equipped with the product [a, b] = ab - ba, is
Lie-isotopic).

We cannot possibly provide a
technical review of the covering genotopic formalism to avoid another
two-semesters volume of lecture notes. However, this presentation would
be insufficient in our view (if not potentially misleading)
without at least the main idea of the genotheories.

In essence, while other scholars searched for {\it departures} from the
Lie axioms, Santilli devoted his research life to {\it preserve} the '
same axioms and search instead for  broader {\it realizations}. This
approach was eventually rewarded because, while other generalizations
outside Lie have the catastrophic physical inconsistencies of Sect. 2,
the preservation of the abstract Lie axioms permitted  their resolutions
while achieving a structurally broader theory.

The main idea of the genoties is best presented in Ref. [12d]
and consists in the identification in  Lie groups and
algebras the following {\it abstract bimodular structure}
$$
A(w) = U\times A(0)\times U^{\dagger} = e^{iX\times w}\times A(0)\times
e^{-iw\times X} = e_>^{iX>w} > A(0) < e_<^{-iw<X}
$$
$$
idA / dw = A\times X - X\times A = A < X - X > A = (A, X),
$$
$$
e_>^{iX>w} = [ e_<^{- i w<X} ]^{\dagger}
\eqno (6.1)
$$
characterized by: I) a modular associative action to the right $>$; II)
a modular-associative axioms to the left $<$; and III) an
inter-relation
between the two actions generally given by Hermitean conjugation.

The genotopic/Lie-admissible formulations
then introduce a {\it realization} of these axioms
more general than that by the isotopc formulations
given by
the mere {\it relaxation of the symmetric character of the isounit}.

This yields {\it two different generalized mathematics},
one for the ordered product to the right $>$
(representing motion forward in time), and one for the
ordered product to the left $<$ (representing motion backward in time),
with {\it two genounits}, {\it two genoproducts}, etc.,
$$
\hat I^> = 1\hat S, A>B = A\times \hat S\times B,
 \hat I^{>}>A = A>\hat I^> = A,
$$
$$
^<\hat I = 1/\hat P, A<B = A\times \hat R\times B,
^<\hat I < A = A<^<\hat I = A
$$
$$
A = A^{\dagger}, B = B^{\dagger}, \hat R = \hat {S}^{\dagger}
\eqno (6.2)
$$

The above elements must then be completed,
for necessary reasons of consistency, with the forward and backward
genofields, genospaces, genodifferential calculus, genogeometries,
etc. [6e,12e].
The explicit Lie-admissible realization of Lie's axioms I, II and III
then reads (at a fixed value of the parameter w, thus without its
ordering)
$$
A(w) =  e_>^{iX>w} > A(0) < e_<^{-iw<X} =
[e^{iX\times S\times w}\times \hat {I}^>]\times S\times
A(0)\times R\times [^<\hat {I}\times e_<^{-iw\times R\times X}]
$$
$$
i \hat d/\hat {d}w = (A, X) = A < X - X > A  = A\times \hat R\times X
- X\times {\hat S}\times A,
$$
$$
X = X^{\dagger}, \hat R = \hat {S}^{\dagger}
\eqno (6.3)
$$

As one can see, the above structures
permit an axiomatic treatment of irreversibility. In fact
the formulation is {\it structurally irreversible} in the sense that it
is
{\it irreversible}  for all possible
conventional,  reversible  Hamiltonians. This is precisely what
needed for a serious study of irreversibility because
all action-at-a-distance interactions are well known to be reversible
while physical reality is irreversible.

The observation
(and admission) of this physical reality is sufficient, alone,
to establish that
{\it irreversibility should be represented with
anything except the Hamiltonian}. Santilli represents irreversibility
with
nonhermitean, thus irreversible, generalized units. The selection of
the units is evidently preferable over an other possible
choices because it assures the invariance of the representation.

In memoirs [5l,6e,12e] Santilli identifies the Lie-admissible structure
of the historical Hamilton equations (those with external terms);
introduces a
new invariant genohamiltonian mechanics; identifies the new
genoquantization; and works out the invariant Lie-admissible
operator theory. These studies have permitted the reduction of the
irreversibility of of our  macroscopic physical reality to
the most elementary levels of nature, such as an electron in the core
of a star considered as external.

Mutatis mutandae, the belief that an electron in the core of
a star is a reversible system or, worse, that it can be described by
quantum mechanics, implies the exiting of science (because it implies
the
belief of the perpetual motion within a hyperdense physical medium
because of the usual
conservation laws of the theory). Rather than adapting
physical reality to [re-existing
theories, Santilli has constructed a theory that can
represent physical reality in an invariant way.

Note that the theory is manifestly
open-nonconservative because $idH/dt =
 (H, H) = H\times (R - S)\times H \not = 0$. Yet, the notion of
{\it genohermiticity} on $\hat {\cal H}^>$ over $\hat C^>$ coincides
with
conventional Hermiticity. Therefore, the Lie-admissible theory provides
the only  operator representation of open systems
known to this author in which
the {\it nonconserved Hamiltonian and other quantities are
Hermitean, thus observable}. In other treatments
of nonconservative systems the Hamiltonian is
generally {\it nonhermitean} and, therefore, {\it not observable}.

Intriguingly, Ref. [12e] proves that the product $A<B - B>A = A\times
{\hat R}\times B - B\times {\hat S}\times A, A\not = B$, is manifestly
non-Lie on conventional spaces over conventional fields, yet it becomes
fully
antisymmetry and Lie when formulates on the bimodule of the
respective envelopes to the left and to the right,
$\{^<\hat {\cal A},\hat {\cal A}^>\}$ (explicitly, the numerical
values of $A\times B$ computed with respect to I is the same
as that of $A>B = A\times {\hat S}\times B$ when computed with
respect to $\hat I^> = 1/\hat S$).

The same quoted contributions on genotopies
identified the limitations of the formulations themselves as being
{\it single-valued} (e.g., a Hamiltonian has only {
\it one} genoeigenvalue per each direction of time).
 Illert and Santilli [13a] provided evidence of the need for
{\it multi-valued methods} in
{\it biological structures}.

In fact,
mathematical treatments complemented with computer visualization
establish that the {\it shape}
of sea shells can be described via the conventional single-valued
three-dimensional Euclidean space according to the empirical perception
of our
three Eustachian tubes. However, the same
space is basically insufficient to represent
{\it the growth in time} of sea shells. In fact,
computer visualization show that, under the exact imposition of the
Euclidean axioms, sea shells first grow in time
in a distorted way and then crack.

Ref. [13a] then showed that the
 minimally consistent representation of sea shells growth requires
{\it six dimensions}. But sea shells exist in our environment and
can be observed via our {\it three-dimensional} perception.
The solution proposed by Santilli [13b] is that
via his {\it multi-valued hypermathematics}  essentially characterized
by
the relaxation of the single-valued character of the genounits
(while preserving their nonsymmetric character as
a necessary condition to represent
irreversible events).

We have in this way the ordered {\it hyperunits} and {\it
 hyperproducts} [6e,13b]
$$
\hat I^> = \{\hat {I}_1^>, \hat {I}_2^>, \hat {I}_3^>, ...\} = 1/\hat
S,
$$
$$
A>B = \{ A\times \hat S_1\times B, A\times \hat S_2\times B,
A\times \hat S_3\times B, ...\},
 \hat I^{>}>A = A>\hat I^> = A,
$$
$$
^<\hat I = \{^<\hat {I}_1, ^<\hat {I}_2, ^<\hat {I}_3, ...\} = 1/\hat S,
$$
$$
A<B = \{A\times \hat R_1\times B, A\times hat R_2\times B,
A\times \hat R_3\times B, ...\}
^<\hat I < A = A<^<\hat I = A
$$
$$
A = A^{\dagger}, B = B^{\dagger}, \hat R = \hat {S}^{\dagger}
\eqno (6.4)
$$

All aspects of the dual Lie-admissible
formalism admit a unique, and significant extension to
the above hyperstructures
(for their expression via{\it  weak equalities and operations }
one may consult Ref. [13c]).

 The belief in the existence of a "final theory for everything" can only
occur to feverish minds because so dissonant with the complexity of
our reality. Despite their remarkable generality,
hyperformulations too cannot describe the entire universe. In fact, as
indicated
in Sect. 2, all available classical theories
(including conventional, isotopic, genotopic and
hyperstructural theories) cannot consistently represent {\it antimatter}
at the {\it classical} level (see the end of Sect. 2).

After several years of research, Santilli resolved the  unbalance
between matter and antimatter
in the physics of this century by introducing (for the first time
in Ref. [5b] of 1985), the map,
called {\it isoduality}, for an arbitrary
quantity A with underlying spaces and fields
$$
A(x, v, \psi, ...)\rightarrow A^d =
-A^{\dagger}(-x^{\dagger}, -v^{\dagger}, -\psi^{\dagger}, ...)
\eqno (6.5)
$$

The above map is mathematically nontrivial, e.g., because it implies the
first
construction on records of {\it numbers with negative units and norm}
[6d].
Physically the map is also nontrivial because it implies an isodual
image of our
universe which coexists with our own, yet it is physically distinct.

We have in this way the {\it isodual conventional, isotopic, genotopic
and
hyperstructural mathematics} [14], which he then applied to the
construction
of a new {\it isodual theory of antimatter}. In particular,
at the operator level,
{\it isoduality is equivalent to charge conjugation}.
The main property here is that
charge conjugation is only applicable at
the level of {\it second quantization},
while {\it isoduality holds for all levels of study, from Newton to
second quantization}, thus resolving the historical unbalance of
this century indicated at the end of Sect. 2.

The reader can see the inevitability of the isodual treatment of
antimatter by noting that {\it the fundamental novel
invariants (3.3) and (3.4) also hold for negative-definite units}.
This guarantees that all properties and physical laws of
the conventional invariants also apply to
antimatter under isoduality, the main difference being that the
treatments of matter and antimatter are anti-isomorphic to each others,
as they should be.

The most general formulation of the theories presented in this paper
is the {\it isoselfdual hypercosmology} [5i],
in which the "universe": has a multi-valued structure perceived by
our Eustachian tubes as a single-valued three-dimensional structure;
is defined to  include
biological structures (as it should be);
is open-irreversible; admits equal amounts of matter and antimatter
(in its
limit formulation verifying Lie's axioms III of Eq. (6.1));
and {\it possesses all identically null total characteristics of
time, energy, linear and angular momentum,
etc.}

In closing, the reader should be aware that isotopic,
genotopic and hyperstructural
formulations and their isoduals can be constructed in their entirety
via simple nonunitary transforms of conventional theories, provided
that they are applied to the {\it totality}
of the  original mathematics. For brevity, we refer the
reader to [5l,10b,12e].

The latter methods are easily applicable for the explicit construction
of
the iso-, geno- and hyper-liftings of the Lorentz-Poincar\'e symmetry,
including the case of nondiagonal-nonsymmetric metrics.
\vskip 0.50 cm

\noindent {\large \bf 7. Concluding remark}

\noindent The question raised by the studies reviewed in this paper is:
why
use generalized theories with limited representational
capabilities and catastrophic physical inconsistencies, rather
when we have available
axiomatically consistent, invariant and universal formulations ?
\vskip 0.50 cm

\noindent {\large \bf Acknowledgments.}

\noindent The author has no words to thank Prof. R. M. Santilli
for invaluable criticisms in the writing of this paper as well as
the use of his computer files.

\end{document}